\documentclass[nojss]{jss}
\usepackage{amsmath}    
\usepackage{amsbsy}
\usepackage{natbib}

\author{Anestis Touloumis\\ Cancer Research UK Cambridge Institute, University of Cambridge}
\title{\proglang{R} Package \pkg{multgee}: A Generalized Estimating Equations Solver for Multinomial Responses}

\Plainauthor{Anestis Touloumis} 
\Plaintitle{R Package multgee: A Generalized Estimating Equations Solver for Multinomial Responses} 
\Shorttitle{\pkg{multgee}: GEE for Multinomial Responses} 

\Abstract{
The \proglang{R} package \pkg{multgee} implements the local odds ratios generalized estimating equations (GEE) approach proposed by \cite{Touloumis2012}, a GEE approach for correlated multinomial responses that circumvents theoretical and practical limitations of the GEE method. A main strength of \pkg{multgee} is that it provides GEE routines for both ordinal (\code{ordLORgee}) and nominal (\code{nomLORgee}) responses, while relevant softwares in \proglang{R} and \proglang{SAS} are restricted to ordinal responses under a marginal cumulative link model specification. In addition, \pkg{multgee} offers a marginal adjacent categories logit model for ordinal responses and a marginal baseline category logit model for nominal. Further, utility functions are available to ease the local odds ratios structure selection (\code{intrinsic.pars}) and to perform a Wald type goodness-of-fit test between two nested GEE models (\code{waldts}). We demonstrate the application of \pkg{multgee} through a clinical trial with clustered ordinal multinomial responses.
}
\Keywords{generalized estimating equations, nominal and ordinal multinomial responses, local odds ratios, \proglang{R}}
\Plainkeywords{generalized estimating equations, nominal and ordinal multinomial responses, local odds ratios, R} 

\Address{
  Anestis Touloumis\\
  Cancer Research UK Cambridge Institute\\
  University of Cambridge\\
  Li Ka Shing Centre, Robinson Way, Cambridge, CB2 0RE, UK\\
  E-mail: \email{Anestis.Touloumis@cruk.cam.ac.uk}
}

\begin{document}

\section{Introduction}
In several studies, the interest lies in drawing inference about the regression parameters of a marginal model for correlated, repeated or clustered multinomial variables with ordinal or nominal response categories while the association structure between the dependent responses is of secondary importance. The lack of a convenient multivariate distribution for multinomial responses and the sensitivity of ordinary maximum likelihood methods to misspecification of the association structure led researchers to modify the GEE method of \cite{Liang1986} in order to account for multinomial responses \citep{Miller1993,Lipsitz1994,Williamson1995,Lumley1996,Heagerty1996,Parsons2006}. These GEE approaches estimate the marginal regression parameter vector by solving the same set of estimating equations as in \cite{Liang1986}, but differ in the way they parametrize and/or estimate $\boldsymbol \alpha$, a parameter vector that is usually defined to describe a ``working'' assumption about the association structure. 

\cite{Touloumis2012} showed that the joint existence of the estimated marginal regression parameter vector and $\hat{\boldsymbol \alpha}$ cannot be assured in existing approaches. This is because the parametric space of the proposed parameterizations of the association structure depends on the marginal model specification even in the simple case of bivariate multinomial responses. To address this issue, \cite{Touloumis2012} defined $\boldsymbol \alpha$ as a ``nuisance'' parameter vector that contains the marginalized local odds ratios structure, that is the local odds ratios as if no covariates were recorded, and they employed the family of association models \citep{Goodman1985} to develop parsimonious and meaningful structures regardless of the response scale. The practical advantage of the local odds ratios GEE approach is that it is applicable to both ordinal and nominal multinomial responses without being restricted by the marginal model specification. Simulations in \cite{Touloumis2012} imply that the local odds ratios GEE approach captures a significant portion of the underlying correlation structure, and compared to the independence `working' model (i.e., assuming no correlation structure in the GEE methodology), simple local odds ratios structures can substantially increase the efficiency gains in estimating the regression vector of the marginal model. Note that low convergence rates for the GEE approach of \cite{Lumley1996} and \cite{Heagerty1996} did not allow the authors to compare these approaches with the local odds ratios GEE approach while the GEE approach of \cite{Parsons2006} was excluded from the simulation design because its use is restricted to a cumulative logit marginal model specification. 

The \proglang{R} \citep{RCoreTeam2013} package \pkg{multgee} implements the local odds ratios GEE approach and it is available from CRAN at \url{http://CRAN.R-project.org/package=multgee}. To emphasize the importance of reflecting the nature of the response scale on the marginal model specification and on the marginalized local odds ratios structure, two core functions are available in \pkg{multgee}: \code{nomLORgee} which is appropriate for GEE analysis of nominal multinomial responses and \code{ordLORgee} which is appropriate for ordinal multinomial responses. In particular, options for the marginal model specification include a baseline category logit model for nominal response categories and a cumulative link model or an adjacent categories logit model for ordinal response categories. In addition, there are three utility functions that enable the user to: i) Perform goodness-of-fit tests between two nested GEE models (\code{waldts}), ii) select the local odds ratios structure based on the rule of thumb discussed in \cite{Touloumis2012} (\code{intrinsic.pars}), and iii) construct a probability table (to be passed in the core functions) that satisfies a desired local odds ratios structure (\code{matrixLOR}).

To appreciate the features of \pkg{multgee}, we briefly review GEE software for multinomial responses in \proglang{SAS} \citep{SAS} and \proglang{R}. The current version of \proglang{SAS} supports only the independence ``working'' model under a marginal cumulative probit or logit model for ordinal multinomial responses. To the best of our knowledge, \proglang{SAS} macros \citep{Williamson1998,Yu2004} implementing the approach of \cite{Williamson1995} are not publicly available. The \proglang{R} package \pkg{repolr} \citep{Parsons2013} implements the approach of \cite{Parsons2006} but it is restricted to using a cumulative logit model. Another option for ordinal responses is the function \code{ordgee} in the \proglang{R} package \pkg{geepack} \citep{Hojsgaard2006}. This function implements the GEE approach of \cite{Heagerty1996} but it seems to produce unreliable results for multinomial responses. To illustrate this, we simulated independent multinomial responses under a cumulative probit model specification with a single time-stationary covariate for each subject and we employed \code{ordgee} to obtain the GEE estimates from the independence `working' model. Description of the generative process can be found in Scenario 1 of \cite{Touloumis2012} except that we used the values $-3,-1,1$ and $3$ for the four category specific intercepts in order to make the problem more evident. Based on $1000$ simulation runs when the sample size $N=500$, we found that the bias of the GEE estimate of $\beta=1$ was $\approx 4.8 \times 10^{28}$, indicating the presence of a bug or -at least- of numerical problems for some situations. Similar problems occurred for the alternative global odds ratios structures in \code{ordgee}. In contrast to existing software, \pkg{multgee} offers greater variety of GEE models for ordinal responses, implements a GEE model for nominal responses and is not limited to the independence ``working'' model, which might lead to significant efficiency losses. Further, one can assess the goodness of fit for two or more nested GEE models.

This paper is organized as follows. In Section \ref{GEENotation}, we present the theoretical background of the local odds ratios GEE approach that is necessary for the use of \pkg{multgee}. We introduce the marginal models implemented in \pkg{multgee}, the estimation procedure for the `nuisance' parameter vector $\boldsymbol \alpha$ and the asymptotic theory on which GEE inference is based. We describe the arguments of the core GEE functions (\code{nomLORgee}, \code{ordLORgee}) in Section \ref{Description1} while the utility functions (\code{waldts}, \code{intrinsic.pars}, \code{matrixLOR}) are described in Section \ref{Description2}. In Section \ref{Example}, we illustrate the use of \pkg{multgee} in a longitudinal study with correlated ordinal multinomial responses. We summarize the features of the package and provide a few practical guidelines in Section \ref{Summary}. 

\section{Local odds ratios GEE approach} \label{GEENotation}
For notational ease, suppose the data arise from a longitudinal study with no missing observations. However, note that the local odds ratios GEE approach is not limited neither to longitudinal studies nor to balanced designs, under the strong assumption that missing observations are missing completely at random \citep{Rubin1976}.

Let $Y_{it}$ be the multinomial response for subject $i$ $(i=1,\ldots,N)$ at time $t$ $(t=1,\ldots,T)$ that takes values in $\{1,2,\ldots,J\}$, $J>2$. Define the response vector for subject $i$
$$\mathbf {Y}_{i}=(Y_{it1},\ldots,Y_{i1(J-1)},Y_{i21},\ldots,Y_{i2(J-1)},\ldots,Y_{iT1},\ldots,Y_{iT(J-1)})^{\top},$$ 
where $Y_{itj}=1$ if the response for subject $i$ at time $t$ falls at category $j$ and $Y_{itj}=0$ otherwise. Denote by $\mathbf{x}_{it}$ the covariates vector associated with $Y_{it}$, and let $\mathbf x_{i}=(\mathbf x^{\top}_{i1},\ldots,\mathbf x^{\top}_{iT})^{\top}$ be the covariates matrix for subject $i$. Define $\pi_{itj}= \E(Y_{itj}|\mathbf x_i)=\Prob(Y_{itj}=1| \mathbf x_i)=\Prob(Y_{it}=j| \mathbf x_i)$ as the probability of the response category $j$ for subject $i$ time $t$, and let $\boldsymbol \pi_{i}=(\boldsymbol \pi^{\top}_{i1},\ldots,\boldsymbol \pi^{\top}_{iT})^{\top}$ be the mean vector of $\mathbf Y_i$, where $\boldsymbol{\pi}_{it} = (\pi_{it1},\ldots,\pi_{it(J-1)})^{\top}$. It follows from the above that $Y_{itJ}=1-\sum_{j=1}^{J-1} Y_{itj}$ and $\pi_{itJ}=1-\sum_{j=1}^{J-1} \pi_{itj}$.

\subsection{Marginal models for correlated multinomial responses}
The choice of the marginal model depends on the nature of the response scale. For ordinal multinomial responses, the family of cumulative link models
\begin{equation}
F^{-1}\left[\Prob(Y_{it}\leq j|\mathbf x_i)\right]=\beta_{0j}+ {\boldsymbol \beta}_{\ast}^{\top} \mathbf{x}_{it}
\label{ABMCLM}
\end{equation}
or the adjacent categories logit model  
\begin{equation}
\log\left(\frac{\pi_{itj}}{\pi_{it(j+1)}} \right)=\beta_{0j}+ {\boldsymbol \beta}_{\ast}^{\top} \mathbf{x}_{it}
\label{ABMACLM}
\end{equation}
can be used, where $F$ is the cumulative distribution function of a continuous distribution and $\{\beta_{0j}:j=1,\ldots,J-1\}$ are the category specific intercepts. For nominal multinomial responses, the baseline category logit model
\begin{equation}
\log\left(\frac{\pi_{itj}}{\pi_{itJ}}\right)=\beta_{0j}+{\boldsymbol {\beta}}_{j}^{\top} \mathbf{x}_{it}
\label{ABMBCLM}
\end{equation}
can be used, where $\boldsymbol {\beta}_{j}$ is the $j$-th category specific parameter vector.

It is worth mentioning that the linear predictor differs in the above marginal models. First, the category specific intercepts need to satisfy a monotonicity condition $\beta_{01}\leq\beta_{02}\leq \ldots \leq \beta_{0(J-1)}$ only when the family of cumulative link models in (\ref{ABMCLM}) is employed. Second, the regression parameter coefficients of the covariates $\mathbf x_{it}$ are category specific only in the marginal baseline category logit model~(\ref{ABMBCLM}) and not in the ordinal marginal models (\ref{ABMCLM}) and (\ref{ABMACLM}).

\subsection{Estimation of the marginal regression parameter vector}
To unify the notation, let $\boldsymbol \beta$ be the $p$-variate parameter vector that includes all the regression parameters in (\ref{ABMCLM}), (\ref{ABMACLM}) or (\ref{ABMBCLM}). To obtain $\boldsymbol {\widehat \beta_G}$, a GEE estimator of $\boldsymbol \beta$, \cite{Touloumis2012} solved the estimating equations
\begin{equation}
\mathbf{U}(\boldsymbol \beta,\widehat{\boldsymbol \alpha})=\frac{1}{N}\sum_{i=1}^N \mathbf{D}_i \mathbf V^{-1}_{i} (\mathbf {Y}_i-\boldsymbol{\pi}_i)=\mathbf{0}
\label{EEbeta}
\end{equation} 
where $\mathbf{D}_i=\partial \boldsymbol{\pi}_i/\partial \boldsymbol{\beta}$ and $\mathbf V_i$ is a $T(J-1) \times T(J-1)$ `weight' matrix that depends on $\boldsymbol \beta$ and on $\widehat{\boldsymbol \alpha}$, an estimate of the `nuisance' parameter vector $\boldsymbol \alpha$ defined formally in Section \ref{Alpha}. Succinctly, $\mathbf V_i$ is a block matrix that mimics the form of $\COV(\mathbf{Y}_i|\mathbf x_i)$, the true covariance matrix for subject $i$. The $t$-th diagonal matrix of $\mathbf V_i$ is the covariance matrix of $Y_{it}$ determined by the marginal model. The $(t,t^{\prime})$-th off-diagonal block matrix describes the marginal pseudo-association of $(Y_{it},Y_{it^{\prime}})$, which is a function of the marginal model and of the pseudo-probabilities $\{\Prob(Y_{it}=j,Y_{it^{\prime}}=j^{\prime}|\mathbf x_i):j,j^{\prime}=1,\ldots,J-1\}$ calculated based on $(\widehat{\boldsymbol \alpha},\boldsymbol \beta)$. We should emphasize that $\mathbf V_i$ is a `weight' matrix because $\boldsymbol \alpha$ is defined as a `nuisance' parameter vector and it is unlikely to describe a valid `working' assumption about the association structure for all subjects.

\subsection{Estimation of the nuisance parameter vector and of the weight matrix} \label{Alpha}
Order the $L=T(T-1)/2$ time-pairs with the rightmost element of the pair most rapidly varying as $(1,2),(1,3),\ldots,(T-1,T)$, and let $G$ be the group variable with levels the $L$ ordered pairs. For each time-pair $(t,t^{\prime})$, ignore the covariates and cross-classify the responses across subjects to form an $J \times J$ contingency table such that the row totals correspond to the observed totals at time $t$ and the column totals to the observed totals at time $t^{\prime}$, and let $\theta_{tjt^{\prime}j^{\prime}}$ be the local odds ratio at the cutpoint $(j,j^{\prime})$ based on the expected frequencies $\{f_{tjt^{\prime}j^{\prime}}:j,j^{\prime}=1,\ldots,J\}$. For notational reasons, let $A$ and $B$ be the row and column variable respectively. Assuming a Poisson sampling scheme to the $L$ sets of $J \times J$ contingency tables, fit the RC-G(1) type model \citep{Becker1989a}
\begin{equation}
\log f_{tjt^{\prime}j^{\prime}}=\lambda+\lambda^{A}_{j}+\lambda^{B}_{j^{\prime}}+\lambda^{G}_{(t,t^{\prime})}+\lambda^{AG}_{j(t,t^{\prime})}+\lambda^{BG}_{j^{\prime}(t,t^{\prime})}+\phi^{(t,t^{\prime})}\mu^{(t,t^{\prime})}_j \mu^{(t,t^{\prime})}_{j^{\prime}}, 
\label{RCGmodel}
\end{equation}
where $\{\mu^{(t,t^{\prime})}_{j}:j=1,\ldots,J\}$ are the score parameters for the $J$ response categories at the time-pair $(t,t^{\prime})$.
After imposing identifiability constraints on the regression parameters in (\ref{RCGmodel}), the log local odds ratios structure is given by
\begin{equation}
\log \theta_{tjt^{\prime}j^{\prime}}=\phi^{(t,t^{\prime})}\left(\mu^{(t,t^{\prime})}_{j}-\mu^{(t,t^{\prime})}_{j+1}\right)\left(\mu^{(t,t^{\prime})}_{j^{\prime}}-\mu^{(t,t^{\prime})}_{j^{\prime}+1}\right).
\label{RCstructure2}
\end{equation}
At each time-pair, (\ref{RCstructure2}) summarizes the local odds ratios structure in terms of the $J$ score parameters and the intrinsic parameter $\phi^{(t,t^{\prime})}$ that measures the average association of the marginalized contingency table. Since the score parameters do not need to be fixed or monotonic, the local odds ratios structure is applicable to both nominal and ordinal multinomial responses.
 
\cite{Touloumis2012} defined $\boldsymbol \alpha$ as the parameter vector that contains the marginalized local odds ratios structure
$$\alpha=\left(\theta_{1121},\ldots,\theta_{1(J-1)2(J-1)},\ldots,\theta_{(T-1)1T1},\ldots,\theta_{(T-1)(J-1)T(J-1)}\right)^{\top}$$
where $\theta_{tjt^{\prime}j^{\prime}}$ satisfy (\ref{RCstructure2}). To increase the parsimony of the local odds ratios structures for ordinal responses, they proposed to use common unit-spaced score parameters $\left(\mu^{(t,t^{\prime})}_{j}=j\right)$ and/or common intrinsic parameters $\left(\phi^{(t,t^{\prime})}=\phi\right)$ across time-pairs. For a nominal response scale, they proposed to apply a homogeneity constraint on the score parameters $\left(\mu^{(t,t^{\prime})}_{j}=\mu_{j}\right)$ and use common intrinsic parameters across time-pairs. To estimate $\boldsymbol \alpha$ maximum likelihood methods are involved by treating the $L$ marginalized contingency tables as independent. Technical details and justification about this estimation procedure can be found in \cite{Touloumis2011a} and \cite{Touloumis2012}.

Conditional on the estimated marginalized local odds ratios structure $\widehat{\boldsymbol \alpha}$ and the marginal model specification at times $t$ and $t^{\prime}$, $\{\Prob(Y_{it}=j,Y_{it^{\prime}}=j^{\prime}|\mathbf x_i):t<t^{\prime},j,j^{\prime}=1,\ldots,J-1\}$ are obtained as the unique solution of the iterative proportional fitting (IPF) procedure \citep{Deming1940}. Hence, $\mathbf V_i$ can be readily calculated and the estimating equations in (\ref{EEbeta}) can be solved with respect to $\boldsymbol \beta$. 

\subsection{Asymptotic properties of the GEE estimator}
Given $\widehat{\boldsymbol \alpha}$, inference about $\boldsymbol \beta$ is based on the fact that $\sqrt{N}(\boldsymbol {\widehat\beta}_G-\boldsymbol \beta)\sim \mathrm{N}(\mathbf 0,\boldsymbol {\Sigma})$ asymptotically,
where 
\begin{equation}
\boldsymbol {\Sigma}=\lim_{N\to\infty} N \boldsymbol {\Sigma}_0^{-1} \boldsymbol {\Sigma}_1 \boldsymbol {\Sigma}_0^{-1},
\label{RobustCovariance}
\end{equation}
$\boldsymbol {\Sigma}_0=\sum_{i=1}^N \mathbf{D}_i^{\top} \mathbf {V}^{-1}_{i} \mathbf{D}_i$ and $\boldsymbol {\Sigma}_1=\sum_{i=1}^N \mathbf{D}_i^{\top} \mathbf {V}^{-1}_{i} \COV(\mathbf{Y}_i|\mathbf x_i) \mathbf {V}^{-1}_{i} \mathbf{D}_i$. For finite sample sizes, $\widehat{\boldsymbol {\Sigma}}$ is estimated by ignoring the limit in (\ref{RobustCovariance}) and replacing $\boldsymbol \beta$ with $\boldsymbol {\widehat \beta}_G$ and $\COV(\mathbf{Y}_i|\mathbf x_i)$ with $(\mathbf {Y}_i-\widehat{\boldsymbol{\pi}}_i)(\mathbf {Y}_i-\widehat{\boldsymbol{\pi}}_i)^{\top}$ in $\boldsymbol {\Sigma}_0$ and $\boldsymbol {\Sigma}_1$. In the literature, $\widehat{\boldsymbol {\Sigma}}/N$ is often termed as ``sandwich'' or ``robust'' covariance matrix of $\boldsymbol {\widehat \beta}_G$.

\section{Description of core functions}\label{Description1}
We describe the arguments of the functions \code{nomLORgee} and \code{ordLORgee}, focusing on the marginal model specification (\code{formula}, \code{link}), data representation (\code{id}, \code{repeated}, \code{data}) and local odds ratios structure specification (\code{LORstr}, \code{LORterm}, \code{homogeneous}, \code{restricted}). For completeness' sake, we also present computational related arguments (\code{LORem}, \code{add}, \code{bstart}, \code{LORgee.control}, \code{ipfp.control}, \code{IM}). The two core functions share the same arguments, except \code{link} and \code{restricted} which are available only in \code{ordLORgee}, and they both create an object of the class \code{LORgee} which admits \code{summary}, \code{coef}, \code{update} and \code{residuals} methods. 

\subsection{Marginal model specification}
For ordinal multinomial responses, the \code{link} argument in the function \code{ordLORgee} specifies which of the marginal models (\ref{ABMCLM}) or (\ref{ABMACLM}) will be fitted. The options \code{\textquotedbl{logit}\textquotedbl}, \code{\textquotedbl{probit}\textquotedbl}, \code{\textquotedbl{cauchit}\textquotedbl} or \code{\textquotedbl{cloglog}\textquotedbl} indicate the corresponding cumulative distribution function $F$ in the cumulative link model (\ref{ABMCLM}), while the option \code{\textquotedbl{acl}\textquotedbl} implies that the adjacent categories logit model (\ref{ABMACLM}) is selected. For nominal multinomial responses, the function \code{nomLORgee} fits the baseline category logit model (\ref{ABMBCLM}), and hence the \code{link} argument is not offered.

The \code{formula} (\code{=response~covariates}) argument identifies the multinomial response variable (\code{response}) and specifies the form of the linear predictor (\code{covariates}), assuming that this includes an intercept term. If required, the $J>2$ observed response categories are sorted in an ascending order and then mapped onto $\{1,2,\ldots,J\}$. To account for a covariate \code{x} with a constrained parameter coefficient fixed to 1 in the linear predictor, the term \code{offset(x)} must be inserted on the right hand side of \code{formula}.

\subsection{Data representation}\label{Data Representation}
The \code{id} argument identifies the $N$ subjects by assigning a unique label to each subject. If required, the observed \code{id} labels are sorted in an ascending order and then relabeled as $1,\ldots,N$, respectively.

The \code{repeated} argument identifies the times at which the multinomial responses are recorded by treating the $T$ unique observed times in the same manner as in \code{id}. The purpose of \code{repeated} is dual: To identify the $T$ distinct time points and to construct the full marginalized contingency table for each time-pair by aggregating the relevant/available responses across subjects. The \code{repeated} argument is optional and it can be safely ignored in balanced designs or in unbalanced designs in which if the $t$-th response is missing for a particular subject then all subsequent responses at times $t^{\prime}>t$ are missing for that subject. Otherwise, it is recommended to provide the \code{repeated} argument in order to ensure proper construction of the full marginalized contingency table. To this end, note that if the measurement occasions are not recorded in a numerical mode, then the user should create \code{repeated} by mapping the $T$ distinct measurement occasions onto the set $\{1,\ldots,T\}$ in such a way that the temporal order of the measurement occasions is preserved. For example, if the measurements occasions are recorded as ``before'', ``baseline'', ``after'', then the levels for \code{repeated} should be coded as $1,2$ and $3$, respectively. 

The dataset is imported via the \code{data} argument in ``long'' format, meaning that each row contains all the information provided by a subject at a given measurement occasion. This implies that \code{data} must include the variables specified in the mandatory arguments \code{formula} and \code{id}, as well as the optional argument \code{repeated} when this is specified by the user. If no \code{data} is provided then the above variables are extracted from the \code{environment} that \code{nomLORgee} and \code{ordLORgee} are called. Currently missing observations, identified by \code{NA} in \code{data}, are ignored.

\subsection{Marginalized local odds ratios structure specification} 
The marginalized local odds ratios is specified via the \code{LORstr} argument. Table~\ref{tab:LOR} displays the structures proposed by \cite{Touloumis2012}. Currently the default option is the time excheangeability structure (\code{\textquotedbl{time.exch}\textquotedbl}) in \code{nomLORgee} and the category excheangeability (\code{\textquotedbl{category.exch}\textquotedbl}) structure in \code{ordLORgee}. The uniform (\code{\textquotedbl{uniform}\textquotedbl}) and category excheangeability structures are not allowed in \code{nomLORgee} because given unit-spaced parameter scores are not meaningful for nominal response categories.

The user can also fit the independence `working' model (\code{LORstr}=\code{\textquotedbl{independence}\textquotedbl}) or even provide the local odds ratios structure (\code{LORstr=\textquotedbl{fixed}\textquotedbl}) using the  \code{LORterm} argument. In this case, an $L \times J^2$ matrix must be constructed such that the $g$-th row contains the vectorized form of a probability table that satisfies the desired local odds ratios structure at the time-pair corresponding to the $g$-th level of $G$. 

\cite{Touloumis2011a} discussed two further versions of the \code{\textquotedbl{time.exch}\textquotedbl} and the RC (\code{\textquotedbl{RC\textquotedbl}}) structures based on using: i) Heterogeneous score parameters (\code{homogeneous}=\code{FALSE}) at each time-pair, and/or ii) monotone score parameters (\code{restricted}=\code{TRUE}), an option applicable only for ordinal response categories. However, it is sensible to employ these additional options only when the local odds ratios structures in Table~\ref{tab:LOR} do not seem adequate.

It is important to mention that the user must provide only the arguments required for the specified local odds ratios structure. For example, the arguments \code{homogeneous}, \code{restricted} and \code{LORterm} are ignored when \code{LORstr=\textquotedbl{uniform}\textquotedbl}. 
\begin{table}
\centering
\begin{tabular}{cccccl}
\hline
\hline
$\log \theta_{tjt^{\prime}j^{\prime}}$ & \code{LORstr}   & Functions & Parameters \\
\hline
$\phi$  & \code{\textquotedbl{uniform}\textquotedbl} & \code{ordLORgee} & 1\\
$\phi^{(t,t^{\prime})}$  &\code{\textquotedbl{category.exch}\textquotedbl} & \code{ordLORgee}  &L\\
$\phi \left(\mu_{j}-\mu_{j+1}\right)\left(\mu_{j^{\prime}}-\mu_{j^{\prime}+1}\right)$ & \code{\textquotedbl{time.exch}\textquotedbl}  & Both  & $J-1$ \\
$\phi^{(t,t^{\prime})}\left(\mu^{(t,t^{\prime})}_{j}-\mu^{(t,t^{\prime})}_{j+1}\right)\left(\mu^{(t,t^{\prime})}_{j^{\prime}}-\mu^{(t,t^{\prime})}_{j^{\prime}+1}\right)$ & \code{\textquotedbl{RC}\textquotedbl}    & Both & $L(J-1)$ \\
\hline
\end{tabular}
\caption{The main options for the marginalized local odds ratios structures in \pkg{multgee}.}
\label{tab:LOR}
\end{table}

\subsection{Computational details} 
The default estimation procedure for the marginalized local odds ratios structure is to fit model (\ref{RCGmodel}) to the full marginalized contingency table (\code{LORem=\textquotedbl{3way}\textquotedbl}) after imposing the desired restrictions on the intrinsic and the score parameters. \cite{Touloumis2011a} noticed that the estimated local odds ratios structure under model (\ref{RCGmodel}) is identical to that obtained by fitting independently a row and columns (RC) effect model \citep{Goodman1985} with homogeneous score parameters to each of the $L$ contingency tables. Motivated by this, an alternative estimation procedure (\code{LORem=\textquotedbl{2way}\textquotedbl}) for estimating the structures \code{\textquotedbl{uniform}\textquotedbl} and \code{\textquotedbl{time.exch}\textquotedbl} was proposed. In particular, one can estimate the single parameter of the \code{\textquotedbl{uniform}\textquotedbl} structure as the average of the $L$ intrinsic parameters $\phi^{(t,t^{\prime})}$ obtained by fitting the linear-by-linear association model \citep{Agresti2002} independently to each of the $L$ marginalized contingency tables. For the \code{\textquotedbl{time.exch}\textquotedbl} structure, one can fit $L$ RC effects models with homogeneous (\code{homogeneous=TRUE})/heterogeneous (\code{homogeneous=FALSE}) score parameters and then estimate the log local odds ratio at each cutpoint $(j,j^{\prime})$ by averaging $\log \hat{\theta}_{tjt^{\prime}j^{\prime}}$ for $t<t^{\prime}$. Regardless of the value of \code{LORem}, the appropriate model for counts is fitted via the function \code{gnm} of the \proglang{R} package \pkg{gnm} \citep{Turner2012}.

In the presence of zero observed counts, a small positive constant can be added (\code{add}) at each cell of the marginalized contingency table to ensure the existence of $\widehat{\boldsymbol \alpha}$. We conjecture that a constant of the magnitude $10^{-4}$ will serve this purpose without affecting the strength of the association structure. 

A Fisher scoring algorithm is employed to solve the estimating equations (\ref{EEbeta}) as in \cite{Lipsitz1994}. The only difference is that now $\widehat{\boldsymbol{\alpha}}$ is not updated. The default way to obtain the initial value for $\boldsymbol \beta$ is via the function \code{vglm} of the R package \pkg{VGAM} \citep{Yee2010}. Alternatively, the initial value can be provided by the user (\code{bstart}). The Fisher scoring algorithm converges when the elementwise maximum relative change in two consecutive estimates of $\boldsymbol \beta$ is less than or equal to a predefined positive constant $\epsilon$. The \code{control} argument controls the related iterative procedure variables and printing options. The default maximum number of iterations is $15$ and the default tolerance is $\epsilon=0.001$.
   
Recall that calculation of the `weight' matrix $\mathbf V_i$ at given values of $(\boldsymbol \beta,\boldsymbol \alpha)$ relies on the IPF procedure. The \code{ipfp.ctrl} argument controls the related variables. The convergence criterion is the maximum of the absolute difference between the fitted and the target row and column marginals. By default, the tolerance of the IPF procedure is $10^{-6}$ with a maximal number of iterations equal to 200.

The \code{IM} argument defines which of the \proglang{R} functions \code{solve}, \code{qr.solve} or \code{cholesky} will be used to invert matrices in the Fisher scoring algorithm. 

\section{Description of utility functions}\label{Description2}
The function \code{waldts} performs a goodness-of-fit test for two nested GEE models based on a Wald test statistic. Let $\mathrm{M_0}$ and $\mathrm{M_1}$ be two nested GEE models with marginal regression parameter vectors $\boldsymbol \beta_0$ and $\boldsymbol \beta_1=(\boldsymbol \beta_0^{\top},\boldsymbol \beta^{\top}_q)^{\top}$, respectively. Define a matrix $\mathbf C$ such that $\mathbf C \boldsymbol \beta_1=\boldsymbol \beta_q$. Here $q$ equals the rank of $\mathbf C$ and the dimension of $\boldsymbol \beta_q$. The hypothesis  
$$H_0: \boldsymbol \beta_q=0 \text{ vs } H_1: \boldsymbol \beta_q \neq 0$$
tests the goodness-of-fit of $\mathrm{M_0}$ versus $\mathrm{M_1}$. Based on a Wald type approach, $H_0$ is rejected at $\alpha \%$ significance level, if $(\mathbf C \widehat{\boldsymbol \beta})^{\top} (N\mathbf C \widehat{\boldsymbol \Sigma}\mathbf C^{\top})^{-1}(\mathbf C \widehat{\boldsymbol \beta}) \geq X_{q}(\alpha)$, where $\widehat{\boldsymbol \beta}$ and $\widehat{\boldsymbol \Sigma}$ are estimated under model $\mathrm{M_1}$ and $X_{q}(\alpha)$ denotes the $\alpha$ upper quantile of a chi-square distribution with $q$ degrees of freedom.

\cite{Touloumis2012} suggested to select the local odds ratios structure by inspecting the range of the $L$ estimated intrinsic parameters under the \code{\textquotedbl{category.exch}\textquotedbl} structure for ordinal responses, or under the \code{\textquotedbl{RC}\textquotedbl} structure for nominal responses. If the estimated intrinsic parameters do not differ much, then the underlying marginalized local odds ratios structure is likely nearly exchangeable across time-pairs. In this case, the simple structures \code{\textquotedbl{uniform}\textquotedbl} or \code{\textquotedbl{time.exch}\textquotedbl} should be preferred because they tend to be as efficient as the more complicated ones. The function \code{intrinsic.pars} gives the estimated intrinsic parameter of each time-pair.

The single-argument function \code{matrixLOR} creates a two-way probability table that satisfies a desired local odds ratios structure. This function aims to ease the construction of the \code{LORterm} argument in the core functions \code{nomLORgee} and \code{ordLORgee}.

\section{Example}\label{Example}
To illustrate the main features of the package \pkg{multgee}, we follow the GEE analysis performed in \cite{Touloumis2012}. The data came from a randomized clinical trial \citep{Lipsitz1994} that aimed to evaluate the effectiveness of the drug Auranofin versus the placebo therapy for the treatment of rheumatoid arthritis. The five-level (1=poor, \ldots, 5=very good) ordinal multinomial response variable was the self-assessment of rheumatoid arthritis recorded at one ($t=1$), three ($t=2$) and five ($t=3$) follow-up months. To acknowledge the ordinal response scale, the marginal cumulative logit model 
\begin{align}
\log \left(\frac{\Prob(Y_{it}\leq j|\mathbf x_i)}{1-\Prob(Y_{it}\leq j|\mathbf x_i)}\right)&=\beta_{0j}+\beta_1 I(time_i=3)+\beta_2 I(time_i=5) +\beta_3 trt_i   \nonumber\\ 
                                                     &+\beta_4 I(b_i=2)+\beta_5 I(b_i=3)+\beta_6 I(b_i=4)+\beta_7 I(b_i=5).
\label{MarginalModelData}
\end{align} 
was fitted, where $i=1,\ldots,301$, $t=1,2,3$, $j=1,2,3,4$ and $I(A)$ is the indicator function for the event $A$. Here $\mathbf x_i$ denotes the covariates matrix for subject $i$ that includes the self-assessment of rheumatoid arthritis at the baseline ($b_i$), the treatment variable ($trt_i$), coded as $(1)$ for the placebo group and $(2)$ for the drug group, and the follow-up time recorded in months ($time_i$).\\
The GEE analysis is performed in two steps. First, we select the marginalized local odds ratios structure by estimating the intrinsic parameters under the \code{\textquotedbl{category.exch}\textquotedbl} structure
\begin{CodeChunk}
\begin{CodeInput}
R> library("multgee")
R> data("arthritis")
R> head(arthritis)
R> intrinsic.pars(y = y, data = arthritis, id = id, repeated = time,
+                 rscale = "ordinal")
\end{CodeInput}
\begin{CodeOutput}
0.6517843 0.9097341 0.9022272
\end{CodeOutput}
\end{CodeChunk}
The range of the estimated intrinsic parameters is small ($\approx 0.26$) which suggests that the underlying marginalized association pattern is nearly constant across time-pairs. Thus we expect the \code{\textquotedbl{uniform}\textquotedbl} structure to capture adequately the underlying correlation pattern. Note that we passed the time variable to the \code{repeated} argument because this numerical variable indicates the measurement occasion at which each observation was recorded.

Now we fit the cumulative logit model (\ref{MarginalModelData}) under the \code{\textquotedbl{uniform}\textquotedbl} via the function \code{ordLORgee} 
\begin{CodeChunk}
\begin{CodeInput}
R> fit <- ordLORgee(formula = y ~ factor(time) + factor(trt) + factor(baseline),
+        link = "logit", id = id, repeated = time, data = arthritis,
+        LORstr = "uniform")
R> summary(fit)
\end{CodeInput}
\begin{CodeOutput}
GEE FOR ORDINAL MULTINOMIAL RESPONSES 
version 1.4 modified 2013-12-01 

Link : Cumulative logit 

Local Odds Ratios:
Structure:         uniform
Model:             3way

call:
ordLORgee(formula = y ~ factor(time) + factor(trt) + factor(baseline), 
    data = arthritis, id = id, repeated = time, link = "logit", 
    LORstr = "uniform")

Summary of residuals:
      Min.    1st Qu.     Median       Mean    3rd Qu.       Max. 
-0.5161000 -0.2399000 -0.0749700  0.0000219 -0.0066990  0.9933000 

Number of Iterations: 5 

Coefficients:
                  Estimate   san.se   san.z Pr(>|san.z|)    
beta01            -1.84315  0.38929 -4.7346      < 2e-16 ***
beta02             0.26692  0.35013  0.7624      0.44585    
beta03             2.23132  0.36625  6.0924      < 2e-16 ***
beta04             4.52542  0.42123 10.7434      < 2e-16 ***
factor(time)3      0.00140  0.12183  0.0115      0.99080    
factor(time)5     -0.36172  0.11395 -3.1743      0.00150 ** 
factor(trt)2      -0.51212  0.16799 -3.0486      0.00230 ** 
factor(baseline)2 -0.66963  0.38036 -1.7605      0.07832 .  
factor(baseline)3 -1.26070  0.35252 -3.5763      0.00035 ***
factor(baseline)4 -2.64373  0.41282 -6.4041      < 2e-16 ***
factor(baseline)5 -3.96613  0.53164 -7.4602      < 2e-16 ***
---
Signif. codes:  0 '***' 0.001 '**' 0.01 '*' 0.05 '.' 0.1 ' ' 1

Local Odds Ratios Estimates:
       [,1]  [,2]  [,3]  [,4]  [,5]  [,6]  [,7]  [,8]  [,9] [,10] [,11] [,12]
 [1,] 0.000 0.000 0.000 0.000 2.257 2.257 2.257 2.257 2.257 2.257 2.257 2.257
 [2,] 0.000 0.000 0.000 0.000 2.257 2.257 2.257 2.257 2.257 2.257 2.257 2.257
 [3,] 0.000 0.000 0.000 0.000 2.257 2.257 2.257 2.257 2.257 2.257 2.257 2.257
 [4,] 0.000 0.000 0.000 0.000 2.257 2.257 2.257 2.257 2.257 2.257 2.257 2.257
 [5,] 2.257 2.257 2.257 2.257 0.000 0.000 0.000 0.000 2.257 2.257 2.257 2.257
 [6,] 2.257 2.257 2.257 2.257 0.000 0.000 0.000 0.000 2.257 2.257 2.257 2.257
 [7,] 2.257 2.257 2.257 2.257 0.000 0.000 0.000 0.000 2.257 2.257 2.257 2.257
 [8,] 2.257 2.257 2.257 2.257 0.000 0.000 0.000 0.000 2.257 2.257 2.257 2.257
 [9,] 2.257 2.257 2.257 2.257 2.257 2.257 2.257 2.257 0.000 0.000 0.000 0.000
[10,] 2.257 2.257 2.257 2.257 2.257 2.257 2.257 2.257 0.000 0.000 0.000 0.000
[11,] 2.257 2.257 2.257 2.257 2.257 2.257 2.257 2.257 0.000 0.000 0.000 0.000
[12,] 2.257 2.257 2.257 2.257 2.257 2.257 2.257 2.257 0.000 0.000 0.000 0.000

pvalue of Null model: <0.0001 
\end{CodeOutput}
\end{CodeChunk}
The \code{summary} method summarizes the fit of the GEE model including the GEE estimates, their estimated standard errors based on the ``sandwich'' covariance matrix and the $p$-values from testing the statistical significance of each regression parameter in (\ref{MarginalModelData}). The estimated marginalized local odds ratios structure can be found in a symmetric $T(J-1) \times T(J-1)$ block matrix written symbolically as
$$\begin{bmatrix}
\begin{array}{cccc}
\mathbf 0                               &\boldsymbol{\Theta}_{12}                 &\ldots  &\boldsymbol{\Theta}_{1T} \\
\boldsymbol{\Theta}_{21}       &\mathbf 0                                &\ldots  &\boldsymbol{\Theta}_{2T} \\
\ldots                                  &\ldots                                   &\ddots  & \ldots          \\
\boldsymbol{\Theta}_{T1}       &\boldsymbol{\Theta}_{T2}        &\ldots  &\mathbf 0
\end{array}
\end{bmatrix}.$$
Each block denotes an $(J-1) \times (J-1)$ matrix. The ($j,j^{\prime}$)-th element of the off-diagonal block $\boldsymbol{\Theta}_{tt^{\prime}}$ represents the estimate of $\theta_{tjt^{\prime}j^{\prime}}$. Based on the properties of the local odds ratios it is easy to see that $\boldsymbol{\Theta}_{tt^{\prime}}=\boldsymbol{\Theta}^{\top}_{t^{\prime}t}$ for $t<t^{\prime}$. Finally, the diagonal blocks are zero to reflect the fact that no local odds ratios are estimated when $t=t^{\prime}$. In our example, $J=5$ and thus each block is a $4 \times 4$ matrix. Since the \code{uniform} structure is selected, all local odds ratios are equal and estimated as $2.257$. Finally, \code{pvalue of Null model} corresponds to the $p$-value of testing the hypothesis that no covariate is significant, i.e., $\beta_1=\beta_2=\beta_3=\beta_4 =\beta_5=\beta_6=\beta_7=0$, based on a Wald test statistic. 

The goodness-of-fit of model (\ref{MarginalModelData}) can be tested by comparing it to a marginal cumulative logit model that additionally contains the age and gender main effects in the linear predictor 
\begin{CodeChunk}
\begin{CodeInput}
R> fit1 <- update(fit, formula = ~. + factor(sex) + age)
R> waldts(fit, fit1)
\end{CodeInput}
\begin{CodeOutput}
Goodness of Fit based on the Wald test 

Model under H_0: y ~ factor(time) + factor(trt) + factor(baseline)
Model under H_1: y ~ factor(time) + factor(trt) + factor(baseline) + factor(sex) + 
    age

Wald Statistic=3.9554, df=2, p-value=0.1384
\end{CodeOutput}
\end{CodeChunk} 

\section{Summary and practical guidelines}\label{Summary}
We described the \proglang{R} package \pkg{multgee} which implements the local odds ratios GEE approach \citep{Touloumis2012} for correlated multinomial responses. Unlike existing GEE softwares, \pkg{multgee} allows GEE models for ordinal (\code{ordLORgee}) and nominal (\code{nomLORgee}) responses. The available local odds ratios structures (\code{LORstr}) in each function respect the nature of the response scale to prevent usage of ordinal local odds ratios structures (e.g., \code{\textquotedbl{uniform}\textquotedbl}) in \code{nomLORgee}. The fitted GEE model is summarized via the \code{summary} method while the estimated regression coefficient can be retrieved via the \code{coef} method. The statistical significance of the regression parameters can be assessed via the function \code{waldts}. A similar strategy to that presented in Section \ref{Example}, can be adopted to analyze GEE models for correlated nominal multinomial responses.

From a practical point of view, we recommend the use of the \code{\textquotedbl{uniform}\textquotedbl} structure for ordinal responses and the \code{\textquotedbl{time.exch}\textquotedbl} structure for nominal especially when the range of the estimated intrinsic parameters (\code{intrinsic.pars}) is small. Based on our experience, some convergence problems might occur as the complexity of the local odds ratios structure increases and/or if the marginalized contingency tables are very sparse. Two possible solutions are either to adopt a simpler local odds ratios structure or to increase slightly the value of the constant added to the marginalized contingency tables (\code{add}). However, we believe that users should refrain from using the independence `working' model unless the aforementioned strategies fail to remedy the convergence problems. To decide on the form of the linear predictor, variable selection model procedures could be incorporated using the function \code{waldts}. 

In future versions of \pkg{multgee}, we plan to permit time-dependent intercepts in the marginal models, to increase the range of the marginal models, by including, for example, the family of continuation-ratio models for ordinal responses, and to offer a function for assessing the proportional odds assumption in models (\ref{ABMCLM}) and (\ref{ABMACLM}).

\bibliographystyle{jss}
\bibliography{jssxxx}
\end{document}